\documentstyle[preprint,prl,aps]{revtex}
\begin{document}
\draft
\flushbottom


\title{On the Statistical Significance of Conductance Quantization}

\author{E.  Bascones\cite{byline}, G. G\'{o}mez-Santos and J.J. S\'{a}enz}

\address{Departamento de F\'{\i}sica de la Materia Condensada and Instituto
Universitario de Ciencia de Materiales ``Nicol\'{a}s Cabrera",
Universidad Aut\'{o}noma de Madrid, 28049 Cantoblanco, Madrid, Spain}
\date{\today}
\maketitle
\begin{abstract}
Recent experiments on atomic-scale metallic contacts have shown that the
quantization of the conductance appears clearly only after the
average of the experimental results.  Motivated by these results we have
analyzed a simplified model system in which a narrow neck is
randomly coupled to wide ideal leads, both in absence and presence of
time reversal invariance.  Based on Random Matrix Theory we study
analytically the probability distribution for the conductance of such
system.  As the width of the leads increases the distribution becomes
sharply peaked close to an integer multiple of the quantum of
conductance.  Our results
suggest a possible statistical origin of conductance quantization
in atomic-scale metallic contacts.
\end {abstract}

The study of electronic transport through mesoscopic structures
has long been a topic of interest. In the last decade much attention has
been focused on this subject in connection with the discovery
of interesting quantum effects that are directly observable
in resistance and conductance experiments. Different phenomena like weak
localization \cite{WL} and universal conductance fluctuations
\cite{UCF1,UCF,Nazarov} appear as a consequence of electron multiple
scattering in the diffusive transport regime. In this regime, the size
of the system is larger then the elastic mean free path $\ell$ but
smaller than the phase-coherence length $L_{\phi}$. When the size of the
system is reduced further, and becomes less than $\ell$,  transport is
said to be ballistic.
The simplest ballistic
system is a short, narrow channel connecting two two-dimensional
electron gases (2DEG).
The conductance of these contacts was shown \cite{CQ} to be quantized in
integer multiples of $2e^2/h$,
even though the ballistic region is strongly coupled to
diffusive structures. This remarkable fact was explained theoretically
by Maslov, Barnes and Kirczenov \cite{Maslov} and Beenakker and Melsen
\cite{Melsen}. The ballistic constriction between two disordered
diffusive leads works as a filter: it  surpresses
fluctuations and recovers conductance quantization (CQ).

The development of different experimental techniques
\cite{Muller,Agrait,Nacho,Olesen,Costa-Kramer,Scheer,Untiedt,Jan}
has made it possible to analize
CQ effects in atomic-scale metallic contacts.
As the size of the contact is of the order of the electron
wave length and much smaller than the elastic mean free path,
they provide a natural system to study mesoscopic transport in three
dimensional structures.
In general, the conductance
of these contacts shows complicated plateau-like structures  that are not
multiples of $2e^2/h$. The
quantization of the conductance manifests itself only after statistical
averaging of different  experimental
realizations.
CQ has been observed as
clear peaks at integer multiples of $2e^2/h$ in the
conductance distribution of
different metallic
contacts.\cite{Muller,Agrait,Nacho,Olesen,Costa-Kramer,Scheer,Untiedt,Jan}
The situation could suggest some kind of
filtering effect similar to that discussed for a 2DEG ballistic
constriction between two diffusive reservoirs.

Motivated by these results we have
analyzed a simplified model system in which a narrow neck
is randomly coupled to wide
ideal leads containing a large number of channels $N$.
Based on  Random Matrix
Theory (RMT)
\cite{UCF,Beenakker-review,Mello,Baranger,chaotic,Bascones}
we will show that conductance quantization is an statistical
property of the system in the large $N$ limit.
We will study analytically
the probability distribution for the conductance of such system and we
will show
that it is sharply peaked near the quantum of conductance. We
stress that our goal here is to  discuss some
general statistical properties of the conductance for a model system
rather than to fit the
experimental results in real atomic-scale contacts.

Our model system is sketched in Fig. 1.
The transport properties are characterized by its
scattering matrix $S$ which is usually written in the form
\[ S =
\left( \begin{array}{clcr}
 r & t' \\
 t & r'
\end{array} \right) ,
\]
where $t$ and $t'$ are the matrices of the transmission amplitudes  for
the waves incident from the left and from the right respectiveley.  The
corresponding matrices of the reflection coeficients are $r$ and $r'$.
Flux conservation implies $S$ unitary while, in the
absence of a magnetic field, time reversal invariance requires $S$ to
be symmetric.
The conductance of the system, (in units of $2 e^{2}/h$)
is given by the two terminal Landauer-B\"{u}ttiker formula
\begin{equation}
g=Tr(tt^{\dagger}) = \sum_{i}T_{i} ,
\end{equation}
where $T_{i}$ are the eigenvalues  of $tt^{\dagger}$.

In our model, we assume a random coupling between the channels in the
narrow constriction and the wide leads (which are assumed to be
perfectly clean).
This is in contrast with the usual approach in the
2DEG context, where a ballistic constriction is connected to
quasi 1D disordered ohmic wires.
As the treatment in terms of RMT makes no reference to
dimensionality, the treatment presented in ref. [6,7],
based on a Transfer matrix approach,
would apply equally well to
3D constrictions connected to ohmic leads.
However, the analysis of CQ in
metallic contacts presents significant differences.

In atomic-scale metallic contacts the randomness could be associated
mainly to the the rough walls of the connecting neck between wide and
narrow leads \cite{volumen}.  It is worth noticing that for metals there
is strong scattering with the constriction walls since the atomic-scale
surface roughness is of the order of the electron wavelength.  The real
situation is even more complicated because of the interplay between
mechanical and electrical
properties.\cite{Agrait,Torres,Landman,Todorov} The experimental peaks
in the histograms are usually obtained after averaging the conductance
of the contact under repeated plastic deformation cycles (in which the
wide leads remain unchanged).  Similar peaks in the conductance
histograms have been obtained in numerical simulations of ballistic
atomic-scale constrictions \cite{Torres} connecting two perfectly clean
seminfinite leads.  In this case the only source of elastic scattering
came from the walls of the neck connecting the perfect leads with the
narrowest section of the contact.  As in the real experiments, the
signature of CQ appears only after averaging over different neck
geometries.\cite{Torres}

Based on these physical considerations, we will consider perfect (non
ohmic) wide leads randomly connected to a narrow constriction.
Since we are interested in a statistical approach, given our limited
knowledge of the microscopic coupling between channels, the natural
choice of the statistical ensemble is that which maximizes the
information
entropy subject to the known constraints.  In the RMT context, this
leads to the circular orthogonal ensemble (COE, $\beta=1$) in the
presence of time-reversal invariance, and to the circular unitary
ensemble (CUE, $\beta=2$) in its absence (for instance, with an applied
magnetic field).\cite{Beenakker-review}

We will first analize the simplest case in which a narrow lead with a
few number of conducting channels $n$ is coupled to a wide one
with $N$ propagating channels (WN geometry) \cite{Stone}.
Following references \cite{Mello,Baranger},
the average ($\langle g \rangle$)
and variance ($var(g) = \langle g^{2} \rangle - \langle g
\rangle
^{2}$) of the conductance can be obtained by averaging on the unitary
group,
\begin{eqnarray}
\langle g \rangle & = & \frac{N n}{N+n+\delta_{1\beta}} \\
var(g) & = &
\frac{
\langle g \rangle^2
(1+\delta_{1\beta})(n+\delta_{1\beta})(N+\delta_{1\beta})}
{(Nn)
(N+n+2\delta_{1\beta}+1)(N+n+\delta_{1\beta}-1)}
\end{eqnarray}
From the last expressions it is obvious that conductance
quantization is recovered in the case that $N>>n$.

A deeper
insight on the statistical properties of $g$ can be obtained
from the
complete distribution of the conductance values $P(g)$.
To compute $P(g)$ we start from the
expression  deduced by Brouwer \cite{Brouwer} for the
distribution of the transmission eigenvalues
\begin{equation}
P(\{T_{p}\})  \propto \prod_{p<m}\mid T_{p} - T_{m} \mid ^{\beta}
\prod_{k}T_{k}^{\alpha} ,
\end{equation}
with $\alpha = \frac{1}{2}\beta(\mid N-n \mid +1-2/\beta)$.
The probability distribution of the conductance can then be obtained from
\begin{equation}
P(g)  =   \int \prod_{i}  dT_{i} \left\{ P(\{T_{i}\})
\delta(g-\sum_i T_i) \right\}.
\end{equation}
Although closed expressions for $P(g)$ can be obtained for small values
of $N$ and $n$ \cite{Baranger}, it is extremely difficult to work it
out for
larger values.
However,
a simple analytical  expression,
for arbitrary $n$,
can be obtained in the large $N$ limit.

Let us write the transmission eigenvalues as $T_k = 1-\epsilon_k$
($\epsilon_k \le 1$).
The conductance can then be written as $g=n-\epsilon$ with
$\epsilon = \sum_k \epsilon_k$.
In the limit $N >> n$ (i.e.  $\alpha >> 1$) we have
\begin{equation}
\prod_{k=1}^{n}T_{k}^{\alpha} \sim e^{-\alpha\epsilon}.
\end{equation}
In terms of $\epsilon$ the probability distribution is now given by
\begin{eqnarray}
P(n-\epsilon)  & = &   e^{-\alpha\epsilon} I(\epsilon) \nonumber \\
I(\epsilon) & = & \int\prod_{k} d\epsilon_{k}\prod_{i<j}
\mid\epsilon_{i}-\epsilon_{j}\mid^{\beta}
\delta(\epsilon-\sum_{m}\epsilon_{m}).
\end{eqnarray}
If $\epsilon<1$ the integration   limits of each variable can be
extended from
zero to infinity
showing that $I(\epsilon)$ is an homogeneous
function of degree $n-1+\beta n(n-1)/2$. Therefore
\begin{equation}
P(g) \propto  e^{-\alpha (n-g)}(n-g)^{n-1+\beta n(n-1)/2}.
\end{equation}
This distribution is the exact large-N  limit in the sense that it
reproduces all
moments of $(n-g)$ to the lowest power   of $(1/N)$.

In Fig. 2 we have  plotted $P(g)$ for different values of n.
The probability  is  peaked at  $g^{*}$
close to integer values but slightly shifted downwards,
\begin{equation}
g^{*}  \sim  n - (n-1) \frac{n \beta/2 + 1}{N \beta/2}.
\end{equation}
As expected, the absence of time reversal invariance (applied magnetic field)
($\beta=2$) surpresses enhanced coherent backscattering
effects, leading to narrower peaks as well as smaller shifts in the peak
positions. Notice also that the quantization `` quality '' of $P(g)$ is
more prominent for larger ratios of $N/n$ (
this is qualitatively similar to the result  obtained within the
Transfer Matrix approach \cite{Melsen}).

Let us know consider the wide-narrow-wide (WNW)
case in which a narrow neck is randomly coupled
to two leads with, in general, different numbers of propagating channels
$N_{1}$ and $N_{2}$ (see Fig. 1).
The conductance of the system is now given by
\begin{equation}
g=Tr[t_{1}t^{\dagger}_{1}(1-r^{\dagger}_{2}r'^{\dagger}_{1})^{-1}t^{\dagger}
_{2}t_{2}(1-r'_{1}r_{2})^{-1}],
\end{equation}
where the $1,2$ index are associated to the coupling with the left and
right lead, respectively.  The direct averaging of this
expression is a formidable task  due to the multiple internal
reflections.
 However, in the limit of $N_{1,2}>>n$ and to lowest order in $1/N_{1,2}$
(i.e. $1/N_{1}$, $1/N_{2}$ or $1/\sqrt{N_{1}N_{2}}$), the expression above
can be
simplified to

\begin{equation}
g=Tr[t_{1}t^{\dagger}_{1} + t^{\dagger}_{2}t_{2} - {\bf 1} +
(r^{\dagger}_{2}r'^{\dagger}_{1} + r'_{1}r_{2})],
\end{equation}
where ${\bf 1}$ is the ($n \times n$) identity matrix.
Assuming statistical independence between left and right leads, it can be
shown\cite{unp} that to leading order, the average and variance of $g$ are
given by
\begin{eqnarray}
\langle g \rangle =
n - n(n+\delta_{1\beta})(\frac{1}{N_{1}}+ \frac{1}{N_{2}}) \\
var(g) = (1+\delta_{1\beta})
n(n+\delta_{1\beta})(\frac{1}{N_{1}}+\frac{1}{N_{2}})^{2}.
\end{eqnarray}
As in the WN case, the previous expressions guarantee that
`` quantization ''  is recovered in the large $N_{1,2}$ limit.  This
means
that the `` filtering '' property of a each WN geometry junction is not
degraded by multiple scattering
between the wide leads.  This intuitively obvious result in
the large $N_{1,2}$ limit can be made rigorous by noticing that the
average conductance of the WNW problem is always greater than $\langle
g_{1} \rangle + \langle g_{2} \rangle - n $, where $g_{1,2}$ are the
conductances of the left and right WN junctions.\cite{unp}

 Now we obtain an approximate probability distribution for the complete
WNW problem, and show it to be equivalent to  classically summing series
resistances. Formally, the approximation consist of ignoring the last
term in the leading order expression for $g$ (Eq. 11) , whose average
 (under the assumption of statistical independence) is
exactly zero. Therefore we take $g=g_1 + g_2 -n$, and the probability
distribution follows
trivially as the   convolution:
\begin{eqnarray}
P(g) = \int dg_{1}  P_{1}(g_{1})  P_{2}(g-(g_{1}-n)),
\end{eqnarray}
where $P_{1,2}(g)$ are the probability distributions (eq. 8) for each WN
contact.
 The previous approach admits a simple physical interpretation. First,
notice that each isolated  WN junction has a conductance $ g_{1,2} = n
- \epsilon_{1,2}$,
 which translates into the following resistance $R_{1,2}= n^{-1} +
\delta R_{1,2} $ with $ \delta R_{1,2} = \epsilon_{1,2}/n $ (remember $
\epsilon_{1,2} = O(1/N_{1,2})$). We can say that $R_{1,2}$ is  the sum
of the perfect
neck resistance $(n^{-1})$ plus an excess {\em contact} resistance
$(\delta R_{1,2})$ which we
can associate with the matching between wide and narrow regions. From a
classical point of view, the complete WNW problen can be seen as the
 series sum of three resistances: the perfect neck resistance
plus the two
contact resistances: $R_{series} = n^{-1} + \delta R_1 + \delta R_2 $,
with corresponding conductance given by $g_{series}
= n - \epsilon_1 - \epsilon_2 = g_1 + g_2 - n $
(again in the large $N_{1,2}$ limit). But $g_{series}$ is precisely the
approximate conductance obtained ignoring the last term of Eq. 11,
leading directly to the distribution of Eq. 14.
This justifies our approximation
as corresponding to the classical sum of series
resistances. Notice that, unlike the WN case, the previous approach does
not reproduce the correct asymptotic large-$N_{1,2}$ distribution. For
instance, it gives the correct average (Eq. 12) but underestimates the
correct variance (Eq. 13). Nevertheless, we expect it to contain
the main qualitative features of the large-$N_{1,2}$ limit. This is
seen, for instance, in Fig. 3 where according to $(14)$
the distribution for the WNW case is plotted for different
cases. As expected, the trends already present in a single WN contact remain
here: decreasing sharpness with increasing $n/N_{1,2}$, and different
behavior related to the presence or absence of enhanced backscattering
($\beta=1,2$).

In conclusion, we have
analyzed a simple model system of a quantum point contact within a
Random $S$ Matrix approach, a choice motivated by experimental
arguments. Within
this model, conductance quantization has an statistical origin.
We have studied analytically the probability distribution for the
conductance.
As a consequence of the random coupling between the narrow and wide
leads,
we have shown that
the conductance distribution  presents peaks at values close to integer
multiples of $2e^2/h$ but slightly shifted to lower values.
The conductance peaks become wider
(the quantization ``quality'' degrades) as the number of channels in the
contact increases.
The absence
of time-reversal invariance (produced by appropiate magnetic fields)
decreases the deviation from quantized values.

The obtained  qualitative behavior (reflected in Figs. 2 and 3)
strikingly resembles
that observed in actual atomic-scale metallic
contacts. The increasing shift and width of the conductance peaks
with  the number of channels in the contact can be clearly seen in the
experimental conductance
histograms\cite{Muller,Agrait,Nacho,Olesen,Costa-Kramer,Scheer,Untiedt,Jan}.
Our results
suggest a possible statistical origin of conductance quantization
in atomic-scale metallic contacts.

We gratefully
acknowledge useful discussions with M.J. Calder\'{o}n and F. Guinea.
Partial support from the DGICyT under Contract No. PB95-0061 and the
EC TMR Project (cod.: FMRX-CT96-0042) is also acknowledged.

\begin{figure}
\caption
{
Sketch of our model system for  the
wide-narrow-wide (WNW) geometry analyzed in the
text.
}
\label{fig1}
\end{figure}

\begin{figure}
\caption
{
Conductance probability distribution $P(g)$ for a WN geometry
for
$N=50$ and different number of channels in the narrow lead
($n=1-4$). Dashed (solid) line corresponds to
$\beta=1$ ($\beta=2$).
}
\label{fig2}
\end{figure}

\begin{figure}
\caption
{
Conductance probability distribution $P(g)$ for a WNW geometry for
$N_{1}=40$,
$N_{2}=60$ and different number  of propagating modes in the neck
($n=1-4$). Dashed (solid) line corresponds to
$\beta=1$ ($\beta=2$).
}
\label{fig3}
\end{figure}


\begin{references}
\bibitem[*]{byline} Present address: Instituto de Ciencia de Materiales, CSIC,
Campus de
Cantoblanco, 28049-Madrid, Spain.

\bibitem{WL} For reviews see :
B.L. Altshuler, A.G. Aronov, D.E. Khmeltnitskii and A.I. Larkin, in
{\em ``Quantum Theory of Solids"}, ed. I.M. Lifshitz (MIR, Moscow,
1982); G. Bergmann, Phys. Rep. {\bf 107}, 1 (1984);
P.A. Lee and T.V. Ramakrishnan, Rev. Mod. Phys. {\bf 57}, 287 (1985);
N.F. Mott and M. Kaveh, Adv. Phys. {\bf 34}, 329 (1985);
S. Chakravarty and A. Schmid, Phys. Rep.{\bf 140}, 193 (1986);
C.W.J. Beenakker and H. van Houten, in {\em Solid State Physics}, {\bf
44}, p. 1, (Academic Press, 1991).

\bibitem{UCF1} C.P. Umbach, S. Washburn, R.B. Laibowitz and R.A. Webb,
Phys. Rev. B {\bf 30}, 4048 (1984);
P.A. Lee, A.D. Stone and H. Fukuyama , {\em ibid.}  {\bf 35}, 1039
(1987).

\bibitem{UCF}
Y. Imry ,
Europhys. Lett {\bf1}, 249 (1986); P.A. Mello, P. Pereyra and N. Kumar,
Ann. Phys. (N.Y.) {\bf 181}, 290 (1988) .P.A. Mello, Phys. Rev. Lett.
{\bf60}, 1089 (1988);
P.A. Mello, E. Akkermans, and B. Shapiro, {\em ibid.} {\bf 61}, 459
(1988).
\bibitem{Nazarov} A. van Oudenaarden et al, Phys. Rev. Lett. {\bf78}, 3539
(1997).
\bibitem{CQ} B.J. Van Wees {\em et al.},
Phys. Rev. Lett. {\bf60}, 848 (1988);
D.A. WHaram {\em et al.},
J. Phys. C {\bf 21}, L209 (1988).

\bibitem{Maslov} D.L. Maslov, C. Barnes, G. Kirczenow,
Phys. Rev. Lett. {\bf70} 1984 (1993);
Phys. Rev. B {\bf48} ,2543 (1993).

\bibitem{Melsen} C.W.J. Beenakker and J.A. Melsen, PRB {\bf50}, 2450  (1994).

\bibitem{Muller} C.J. Muller, J.M. van Ruitenbeek, and L.J. Jongh,
Phys. Rev. Lett.
{\bf69}, 140 (1992);
J.M. Krans {\em et al.}, {\em ibid.} {\bf 74}, 2146 (1995);
Nature (London) {\bf 375}, 6534 (1995).
\bibitem{Agrait} N. Agrait, J.G. Rodrigo and S. Vieira,
Phys. Rev. B {\bf47}, 12345 (1993); G. Rubio, N. Agrait and S. Vieira,
Phys. Rev. Lett. {\bf 76}, 2302 (1996).
\bibitem{Nacho}
J.I. Pascual {\em et al.}, Phys. Rev. Lett.
{\bf 71}, 1852 (1993); Science {\bf 267}, 1793 (1995).
\bibitem{Olesen} L. Olesen {\em et al.}, Phys. Rev. Lett.
{\bf 72}, 2251 (1994); {\em ibid.} {\bf 74}, 2147 (1995);
M. Brandbyge {\em et al.}, Phys. Rev. B
{\bf52}, 8499 (1995).
\bibitem{Costa-Kramer} J.L. Costa-Kramer {\em et al.}
Surf. Sci. Lett.   {\bf342}, L1144 (1995); Phys. Rev. B {\bf 55}, 5416
(1997).
\bibitem{Scheer} E. Scheer et al, Phys. Rev. Lett. {\bf 78}, 3535 (1997).
\bibitem{Untiedt} C. Untiedt et al, Phys. Rev. B {\bf 56}, 2154
(1997).
\bibitem{Jan} For a recent review see: J.M. van Ruitenbeek, in {\em
Mesoscopic electron transport}, NATO-ASI Series {\em in press}, (Kluwer,
Dordrecht, 1997).
\bibitem{Beenakker-review}  C.W.J. Beenakker, {\em `` Random
Matrix
Theory of Quantum Transport"}, Rev. Mod. Phys., in press (1997);
M.L. Mehta,
{\em `` Random Matrices"} (Academic Press, New York,1991).

\bibitem{Mello} P.A. Mello, J.Phys. A. {\bf23}, 4061  (1990).

\bibitem{Baranger} H.U. Baranger and P.A. Mello,
Phys. Rev. Lett. {\bf73}, 142 (1994);
Phys. Rev. B {\bf54}, R14297 (1996);
R.A. Jalabert, J.L. Pichard and C.W.J. Beenakker, Europhys. Lett. {\bf27},
255 (1994).

\bibitem{chaotic}
R.A. Jalabert, A. D. Stone, and Y. Alhassid, Phys. Rev. Lett. {\bf23}, 3468 (199
2); V.A. Gopar, P.A. Mello and M. B\"{u}ttiker, {\em ibid.} {\bf77}, 3005 (1996)
;
P.W. Brouwer
and M. B\"{u}ttiker, Europhys. Lett. {\bf37}, 442 (1997).

\bibitem{Bascones} E. Bascones {\em et al.}, Phys. Rev. B {\bf 55},
R11911 (1997).

\bibitem{volumen}
Volume defects inside the contact region could also be present. However,
Molecular Dynamics simulations predict the existence of crystalline
order within an atomic-scale contact\cite{Landman}.
Moreover, Point Contact Spectroscopy
measurements \cite{Untiedt} indicate that metallic constrictions can be
quite crystalline, showing an almost perfect ballistic
transport
even for conductances as large as hundreds of quanta.

\bibitem{Landman} U. Landman {\em et al.},
Science {\bf 248}, 454 (1990); Phys. Rev. Lett. {\bf 77}, 1362 (1996)

\bibitem{Todorov}
T.N. Todorov and A.P. Sutton, Phys. Rev.
Lett. {\bf 70}, 2138 (1993); Phys. Rev. B {\bf 54}, R14234 (1996).

\bibitem{Torres} J.A. Torres and J.J. S\'{a}enz, Phys. Rev. Lett. {\bf 77},
2245 (1996).


\bibitem{Stone}
WN and WNW geometries were analized by
A. Szafer and A.D. Stone, [Phys. Rev. Lett {\bf 62}, 300 (1989)]
in the context of ballistic transport,

\bibitem{Brouwer} P.W. Brouwer, as quoted by Beenakker in ref. [16].

\bibitem{unp} G. G\'{o}mez-Santos and E. Bascones, unpublished.
\end{references}
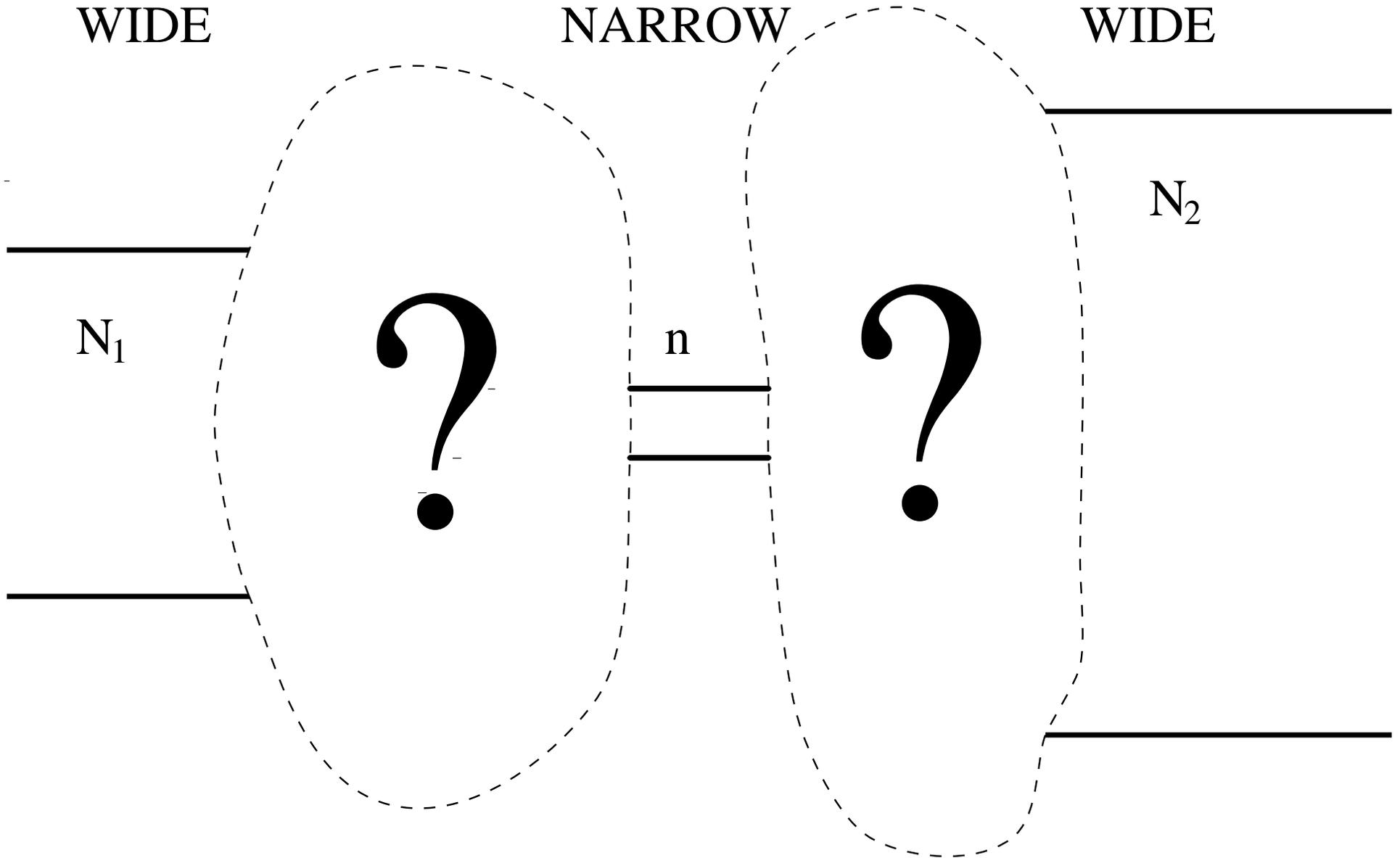
\end{document}